\documentclass{article}

\usepackage{epsfig} 
\usepackage{amsmath}
\usepackage{amsfonts}
\usepackage{graphicx}

 \usepackage{amssymb}

\date{\today}

\newcommand{\insertplot}[5]{\begin{figure}
 \hfill\hbox to 0.05in{\vbox to #5in{\vfill
 \inputplot{#1}{#4}{#5}}\hfill}
 \hfill\vspace{-.1in}
 \caption{#2}\label{#3}
 \end{figure}}
 \newcommand{\inputplot}[3]{
 \special{ps: plotfile #1}
}
\newcounter{fig}

\begin{document}

\title{ 
An instability of the Reissner-Nordstr\"om solution and 
\\
 new hairy black holes in $d=5$ dimensions
} 
  
\author{{\large Yves Brihaye,}$^{\dagger}$
{\large Eugen Radu}$^{\ddagger}$
and {\large D. H. Tchrakian}$^{\star \diamond }$ \\  
$^{\dagger}${\small Physique-Math\'ematique, Universit\'e de Mons, Mons, Belgium}\\
$^{\ddagger}${\small  Institut f\"ur Physik, Universit\"at Oldenburg, Postfach 2503
D-26111 Oldenburg, Germany}
  \\
$^{\star}${\small  Department of Computer Science,
National University of Ireland Maynooth,
Maynooth,
Ireland} \\
$^{\diamond}${\small School of Theoretical Physics -- DIAS, 10 Burlington
Road, Dublin 4, Ireland }}

\newcommand{\ta}{\theta}
\newcommand{\Si}{\Sigma}
\newcommand{\vf}{\varphi}
\newcommand{\dd}{\mbox{d}}
\newcommand{\tr}{\mbox{tr}}
\newcommand{\la}{\lambda}
\newcommand{\ka}{\kappa}
\newcommand{\f}{\phi}
\newcommand{\al}{\alpha}
\newcommand{\ga}{\gamma}
\newcommand{\de}{\delta}
\newcommand{\si}{\sigma}
\newcommand{\bomega}{\mbox{\boldmath $\omega$}}
\newcommand{\bsi}{\mbox{\boldmath $\sigma$}}
\newcommand{\bchi}{\mbox{\boldmath $\chi$}}
\newcommand{\bal}{\mbox{\boldmath $\alpha$}}
\newcommand{\bpsi}{\mbox{\boldmath $\psi$}}
\newcommand{\brho}{\mbox{\boldmath $\varrho$}}
\newcommand{\beps}{\mbox{\boldmath $\varepsilon$}}
\newcommand{\bxi}{\mbox{\boldmath $\xi$}}
\newcommand{\bbeta}{\mbox{\boldmath $\beta$}}
\newcommand{\ee}{\end{equation}}
\newcommand{\eea}{\end{eqnarray}}
\newcommand{\be}{\begin{equation}}
\newcommand{\bea}{\begin{eqnarray}}

\newcommand{\ii}{\mbox{i}}
\newcommand{\e}{\mbox{e}}
\newcommand{\pa}{\partial}
\newcommand{\Om}{\Omega}
\newcommand{\vep}{\varepsilon}
\newcommand{\bfph}{{\bf \phi}}
\newcommand{\lm}{\lambda}
\def\theequation{\arabic{equation}}
\renewcommand{\thefootnote}{\fnsymbol{footnote}}
\newcommand{\re}[1]{(\ref{#1})}
\newcommand{\R}{{\rm I \hspace{-0.52ex} R}}
\newcommand{\N}{{\sf N\hspace*{-1.0ex}\rule{0.15ex}%
{1.3ex}\hspace*{1.0ex}}}
\newcommand{\Q}{{\sf Q\hspace*{-1.1ex}\rule{0.15ex}%
{1.5ex}\hspace*{1.1ex}}}
\newcommand{\C}{{\sf C\hspace*{-0.9ex}\rule{0.15ex}%
{1.3ex}\hspace*{0.9ex}}}
\newcommand{\eins}{1\hspace{-0.56ex}{\rm I}}
\renewcommand{\thefootnote}{\arabic{footnote}}

\def\theequation{\thesection.\arabic{equation}}

\maketitle

\begin{abstract}

The $d=5$ Reissner-Nordstr\"om black hole becomes unstable
when considered as a solution of Einstein--Yang-Mills--Chern-Simons theory.
The existence of a marginally stable mode
  gives rise to a new branch of
 black holes with non-Abelian magnetic fields outside the horizon.
These solutions carry a nonzero electric charge and have finite mass.
We argue that these features manifest
themselves both for Minkowski and (Anti-)de Sitter asymptotics.
The properties of 
solutions in a de Sitter background   are new to the present work and 
are emphasised.  
All solutions constructed have finite mass by virtue of the presence of the
Chern-Simons term, which in $d=5$ plays the role of a higher order term scaling appropriately
for this purpose. 
  
\end{abstract}
\medskip

%%%%%%%%%%%%%%%%%%%%%%%%%%%%%%%%%%%%%%%%%%%%%%
\section{Introduction} 
%%%%%%%%%%%%%%%%%%%%%%%%%%%%%%%%%%%%%%%%%%%%%
The Reissner-Nordstr\"om (RN) black hole is perhaps the 
simplest and the best known non-vacuum configuration in general relativity.
This solution
of the Einstein-Maxwell equations describes the gravitational field of 
an electrically charged\footnote{In $d=4$
spacetime dimensions,
 due to the electric-magnetic
 duality, one may consider as well a
 magnetic charge source for this solution.}, spherically symmetric body.
Discovered very soon  \cite{Reissner,Nord} after the Schwarzschild black hole, the RN solution
enjoyed constant interest since then, being studied from different directions.
This includes solutions in the presence of  a cosmological constant \cite{Kottler}, and,
 generalizations to spacetime dimensions higher than $d=4$ \cite{Myers:1986un}.

The stability of the RN black hole is an important issue.
As discussed in \cite{Moncrief:1974gw}, \cite{Moncrief:1974ng},
the $d=4$ solution is stable with respect to linear perturbations;
the stability of higher dimensional solutions has  been considered in 
\cite{Konoplya:2007jv}, \cite{Ishibashi:2011ws}.
 Moreover, a RN black hole is thermally stable
for large enough values of the electric charge \cite{Davies:1978mf}.
 
There is  however a different type of instability that the RN solution  
can  exhibit. 
This occurs when the Einstein-Maxwell (EM) system is considered 
as part of a larger theory, typically containing scalar fields.  
Then, the RN black hole can become unstable to forming hair at low temperatures.
This results in new branches of hairy solutions of the
larger theory, with the extra matter fields in a non-vacuum state.
Usually, the hairy black holes are thermodynamically favoured over the RN solution.

Although   this type of instability was  originally discovered for $d=4$ asymptotically flat,
magnetically charged, RN black holes embedded in Einstein-Yang-Mills-Higgs theory \cite{Lee:1991qs},
  it has recently received a considerable
  attention in the case of black holes with anti-de Sitter (AdS) asymptotics.
The main models considered in this context are the gravitating Abelian Higgs \cite{Gubser:2008px}
and Einstein-Yang-Mills \cite{Gubser:2008zu}, \cite{Gubser:2008wv} theories,
augmented with a negative cosmological term. 
The resulting sets of hairy black holes possess Ricci-flat horizons, and
have found interesting applications in the context of gravity/gauge duality,
providing a dual gravitational description of some superconductors  (see $e.g.$  \cite{Cadoni:2011kv}, \cite{Abdalla:2010nq};
for a review of this domain, see
\cite{Horowitz:2010gk}).

The main purpose of this work is to discuss a 
a specific type of instability of the electrically charged RN black holes in $d=5$
spacetime dimensions in context of the Einstein--Yang-Mills--Chern-Simons (EYMCS) theory.
The presence of a cosmological term, positive or negative, is immaterial in the sense that we
demonstrate qualitatively similar features in all these cases. This unified treatment of the
various asymptotics is part of our brief here.

Usually, the RN solution is written in Scwarzschild-like coordinates, with a line element
\bea
\label{RN-metric}
ds^2=\frac{dr^2}{N(r)}+r^2d\Omega_3^2-N(r)dt^2, 
\eea
with 
\bea
\label{RN-N}
N(r)= 1-\frac{m(r)}{r^2}+\varepsilon \frac{r^2}{L^2},~~m(r)=M_0-\frac{\alpha^2 Q^2}{r^2},
\eea
(where $\Lambda=6 \varepsilon /L^2$ is the cosmological constant, with $\varepsilon =-1,0,+1$
for AdS, Minkowski and dS asymptotics, respectively, and $d\Omega^2_{3}$ 
is the line element of the three dimensional sphere), and an $U(1)$ electric potential 
\bea
\label{RN-U(1)}
V(r)=V_0-\frac{Q}{r^2}.
\eea
$M_0$ and $Q$ corresponds (up to some constant factors) to the mass and electric charges of
the solution; also we note that $\alpha^2=16\pi G/(3e^2)$, with $G$ the Newton constant and
$e$   is the non-Abelian (nA)  gauge coupling constant.
 
The instability discussed in this work occurs when embedding 
the RN solution (\ref{RN-metric})-(\ref{RN-U(1)}) in 
EYMCS theory, the Chern-Simons (CS) term in which plays a crucial role. 
In the absence of a cosmological constant, this model has been discussed in \cite{Brihaye:2010wp},
\cite{Brihaye:2011nr}, where the corresponding zero mode 
of the RN black hole was exhibited, together with the resulting
set of hairy black hole solutions. The results there indicate 
the existence of a  phase transition between RN solutions 
and the non-Abelian black holes, which generically are thermodynamically preferred.  
Solutions of the EYMCS  model with a negative cosmological constant 
were discussed in  \cite{Brihaye:2009cc}.

One objective here is to argue that the instability of the RN black hole discovered in
\cite{Brihaye:2010wp,Brihaye:2011nr}, \cite{Brihaye:2009cc} for Minkowskian and AdS asymptotics respectively,
occurs also in the presence of a positive cosmological constant in the action.  
As a result,
a unified picture of this instability manifests itself in these three cases. 
For a fixed value of the CS coupling constant, a classical instability is developed
if the horizon is sufficiently small. 
We find that a number of basic features of the resulting set of hairy black
holes are the same both for asymptotically flat, and, for (anti)-de Sitter solutions.
Some of the EYMCS solutions with $\Lambda<0$
in \cite{Brihaye:2009cc} are interpreted here in this context.
Finite mass solutions with dS asymptotics are also constructed.
In all cases, an Abelian gauge symmetry is spontaneously broken near a black hole horizon
with the appearance of a condensate of non-Abelian (nA) gauge fields there.  

The paper is organized as follows. In the next Section we define the model
and propose a spherically symmetric Ansatz.
Section 3 is devoted to the stability analysis of the RN solution in EYMCS theory.
We derive the pulsation equation governing the evolution of radial
nA perturbations and solve it numerically.
Special emphasis is placed on the issue of marginally stable modes, 
an analytic expression being found in this case.
In Section 4 we
 present  numerical results and discuss qualitative properties
of the hairy black hole solutions.
For solutions with a positive cosmological constant, 
we construct new sets of black holes with nA hair approaching
asymptotically the dS background.
We conclude with Section 5, where the significance of, 
and further consequences arising from, the solutions
we have constructed are discussed.

%%%%%%%%%%%%%%%%%%%%%%%%%%%%%%%%%%%%%%%%%%%%%%%%%%%%%%%%%%%%%%%%%%
\section{The  model }
%%%%%%%%%%%%%%%%%%%%%%%%%%%%%%%%%%%%%%%%%%%%%%%%%%%%%%%%%%%%%%%%% 
%%%%%%%%%%%%%%%%%%%%%%%%%%%%%%%%%%%%%%%%%%%%%%%%%%%%%%%%%%%%%%%%%%
\subsection{Field equations }
%%%%%%%%%%%%%%%%%%%%%%%%%%%%%%%%%%%%%%%%%%%%%%%%%%%%%%%%%%%%%%%%%
We consider the following EYMCS action 
\bea
\label{action}
S= \int_{ \mathcal{M}}  d^5x \sqrt{-g} \left( \frac{1}{16\pi G}
 (R-2 \Lambda)-\frac14\,\mbox{Tr}\{ {\cal F}_{\mu\nu}{\cal F}^{\mu\nu} \}  \right)
-\int_{ \mathcal{M}}  d^5x~{\cal L}_{\rm{CS}} ,
\eea
%where 
 %is the usual Yang-Mills lagrangian for a gauge group $SO(6)$ 
(with ${\cal F}_{\mu\nu}=\partial_\mu {\cal A}_\nu-\partial_\nu {\cal A}_\mu+e[{\cal A}_\mu,{\cal A}_\nu]$ the gauge field
strength tensor),
and 
\begin{eqnarray}
&&{\cal L}_{\rm{CS}} =  \kappa \,
\vep_{\mu\nu\rho\si\tau}\mbox{Tr}\, \bigg \{
{\cal A}^{\tau}\left[{\cal F}^{\mu\nu}{\cal F}^{\rho\si}-
e {\cal F}^{\mu\nu} {\cal A}^{\rho}{\cal A}^{\si}+
 \frac25 e^2 {\cal A}^{\mu}{\cal A}^{\nu}{\cal A}^{\rho}{\cal A}^{\si}\right]\, \bigg \},
\label{CS5}
\end{eqnarray}
is the CS term.
The field equations are found by taking the variation of (\ref{action}) 
with respect to $g_{\mu \nu},{\cal A}_\mu$
and are given $e.g.$ in \cite{Brihaye:2009cc}.

There is some arbitrariness in the choice of the gauge group,
the only restriction being that it should be large
 enough to accomodate
%for a static spherically symmetric Ansatz, with
a nonvanishing electric potential, such that the RN black hole is a solution of the model.
  The electric potential is supported by the CS term, which would likewise vanish
if the gauge group is not large enough. Subject to these criteria, 
%Under these assumptions,
the smallest simple gauge group supporting a nonvanishing CS term\footnote{Then, for $\Lambda<0$
this model can be viewed as a truncation of the five-dimensional maximally gauged supergravity
\cite{Gunaydin:1985cu}, \cite{Cvetic:2000nc}, which plays a prominent role in the conjectured
AdS/CFT correspondence \cite{Maldacena:1997re}, \cite{Witten:1998qj}.
One expects  that the basic properties of the solutions in this truncation, persist in the full
model.} is $SO(6)$.

In this work,  
as in \cite{Brihaye:2011nr}, we shall restrict to a $SO(4)\times SO(2)$ truncation of the
$SO(6)$ gauge group, $i.e.$
\begin{eqnarray}
\label{YMansatz}
{\cal A}=\vec A\cdot d\vec r+a dt {\bf T},
\end{eqnarray}
where the magnetic potential $\vec A$
takes values in the $SO(4)\sim SU(2)\times SU(2)$ subgroup, while ${\bf T}$ is the Casmir operator corresponding to the generator of the $SO(2)$ subgroup. 

For such a truncation of the gauge group, \re{YMansatz}, 
it is important to note that the  EYMCS  model admits an equivalent formulation. 
Restricting to an $SU(2)$ subgroup of $SO(4)$, and
discarding a boundary term   arising  in the CS term,
one finds that (\ref{action}) reduces essentially to
a Einstein--Yang-Mills--Maxwell model, with a Chern-Simons-type coupling term between
the $U(1)$ and nA fields
\begin{eqnarray}
\label{action1}
S= \int_{ \mathcal{M}}  d^5x \sqrt{-g} \left( \frac{1}{16\pi G} (R-2 \Lambda)-\frac{1}{4}F_{\mu \nu}^{I}F^{\mu \nu I} 
-\frac{1}{4}f_{\mu \nu} f^{\mu \nu}-
\frac{\kappa }{\sqrt{-g}}\epsilon^{\mu \nu\rho\sigma\tau}a_{\mu}F_{\nu \rho}^{I}F_{\sigma \tau}^{I}
 \right),
%-\kappa \int_{ \mathcal{M}}  d^5x~\epsilon^{\mu \nu\rho\sigma\tau}a_{\mu}F_{\nu \rho}^{I}F_{\sigma \tau}^{I} .
\end{eqnarray}
where $A_{\mu}^I$ is the $SU(2)$ non-Abelian gauge field 
(with $I=1,2,3$ and field strength $F_{\mu\nu}^I=\partial_{\mu}A_{\nu}^{I}-\partial_{\nu}A_{\mu}^{I}+ e \epsilon_{IJK}A_{\mu}^{J}A_{\nu}^{K}$),
and  $a_\mu$ is
the $U(1)$ gauge field ($f_{\mu\nu}=\partial_{\mu}a_{\nu} -\partial_{\nu}a_{\mu}$  being the corresponding field strength).
 
Variation of (\ref{action1})
with respect to $g_{\mu\nu}$, $A_\mu^I$ and $a_\mu$
leads to the following equations:
\begin{eqnarray}
\nonumber
&&
R_{\mu \nu}-\frac{1}{2}R g_{\mu \nu}+\Lambda g_{\mu\nu}=8 \pi G  T_{\mu \nu},
\\
&&
\label{field-eqs}
 \frac{\partial}{\partial x^\mu}
 \left(
 \sqrt{-g} f^{\mu\nu}
 \right)=
\kappa \epsilon^{\nu\alpha \beta \gamma \mu }F^{\alpha \beta I} F^{\gamma \mu I},
\\
&&
\nonumber
\frac{\partial}{\partial x^\mu }
 \left(
 \sqrt{-g} F^{\mu\nu I} \right)
+e \sqrt{-g} \epsilon ^{IJK}A_\mu ^J F^{\mu \nu K}=
 2\kappa \epsilon^{\nu\alpha \beta \gamma \mu }
 f^{\alpha \beta }F^{\gamma \mu I},
\end{eqnarray}
where
\begin{eqnarray}
\label{Tij}
T_{\mu \nu}=F_{\mu\alpha}^I F_{\nu\beta}^I g^{\alpha\beta}-\frac{1}{4}g_{\mu\nu}F_{\alpha\beta}^I F^{\alpha\beta I}
+
f_{\mu\alpha}f_{\nu\beta}g^{\alpha\beta}-\frac{1}{4}g_{\mu\nu}f_{\alpha\beta} f^{\alpha\beta }
 \end{eqnarray}
is the energy-momentum tensor.
Some features of this model with $\Lambda=0$ have been discussed in \cite{Gibbons:1993xt},
where it was found that the value 
$\kappa=\alpha/8$ (with $\alpha^2=16\pi G/(3e^2)$) is special,
being  given by the coupling of the super-YM theory 
to the $d = 5$ supergravity.
This allows to promote all $d=4$ flat space instantons to extremal black hole and soliton
solutions of the Eqs. (\ref{field-eqs}) with a vanishing cosmological constant,
see the discussion  in \cite{Brihaye:2011nr}.
The model (\ref{action1}) can also be thought 
as a truncation with a 
vanishing dilaton of the
${\cal{N}}=4,~d=5$  Romans' gauged supergravity model \cite{Romans:1985ps}.

%%%%%%%%%%%%%%%%%%%%%%%%%%%%%%%%%%%%%%%%%%%%%%%%%%%%%%%%%%%%%%%%%%
\subsection{A spherically symmetric ansatz and the Abelian solutions}
%%%%%%%%%%%%%%%%%%%%%%%%%%%%%%%%%%%%%%%%%%%%%%%%%%%%%%%%%%%%%%%%%

In this work we are interested in spherically symmetric solutions only,
 with  a line element
\begin{eqnarray}
\label{metric-gen}  
ds^{2}=-  N\sigma^2 dt^{2}+\frac{dr^2}{N} +r^2 d\Omega^2_{3}, 
\end{eqnarray} 
where $N,\sigma$ are functions of  $ r$ and $t$ in general.
The gauge fields ansatz contains only two functions, a magnetic $SU(2)$ gauge potential
$w$, and an $U(1)$ electric one, $V$:
\begin{eqnarray}
\label{ansatz-gf}
A^I=\frac{1}{e}(1+w(r,t)) \theta^I,~~a=\frac{1}{e}V(r,t)dt,
\end{eqnarray}
where $\theta^I$ are the invariant forms on $S^3$,
such that $d\theta^I=\epsilon^I_{JK}\theta^J\land \theta^k$ and 
$d\Omega^2_3=\theta^I\theta^I$.

The RN Abelian solution is recovered for   $N$ given by (\ref{RN-N}), $\sigma=1$,
 $w=\pm 1$, and $V$ given by (\ref{RN-U(1)}).
This configuration has a black hole event horizon, located at $r=r_h$ (which is the largest root of the equation $N(r)=0$
satisfying the condition $N'(r_h)>0$). 
The conserved quantities associated
with the RN solution are the mass ${\cal M}=\frac{3\pi}{8 G}M$ and the
electric  charge ${\cal Q}=\frac{4 \pi^2 }{e}Q$.
Other quantities of interest are the chemical potential $\Phi=\frac{V_0}{e}$,
the Hawking temperature $T_H= \frac{1}{2 \pi r_h}(1-\frac{\alpha^2 Q^2 }{r_h^4}+2 \varepsilon \frac{r_h^2}{L^2})$, 
and the entropy $S=\frac{A_H}{4G}$ (with $A_H=2\pi^2 r_h^3$ the black hole event horizon area).
 
 For the discussion of the solutions, it is useful to introduce the dimensionless quantities
\begin{eqnarray}
\label{def-U}
 U=\frac{Q}{r_h^2}=\frac{(2\pi^2)^{2/3}Q}{A_H^{2/3}},~~{\rm and}~~Y= \frac{r_h^2}{L^2}=\frac{A_H^{2/3}}{(2\pi^2)^{2/3}L^2 }.
\end{eqnarray}
 These parameters characterize  uniquelly the RN black hole (asymptotically flat solutions have $Y=0$).
For example,   several dimensionless quantities of interest in the thermodynamical
description of the system can be expressed in terms of $U,Y$:
\begin{eqnarray}
\label{funct-U1} 
f=\frac{F}{Q}=\frac{{\cal M}-T_H S}{Q}=\frac{2\pi^2}{3U}(1+5U^2-Y),~~t_H= {T_H}{Q}^{1/2}=\frac{\sqrt{U}}{2\pi}(1-U^2+2\varepsilon Y),
~~a_H=\frac{A_H}{Q^{3/2}}=\frac{2\pi^2}{U^{3/2}},~~
\end{eqnarray}
where $f,~t_H$ and $a_H$ are the reduced free energy, temperature and
area, respectively.

%%%%%%%%%%%%%%%%%%%%%%%%%%%%%%%%%%%%%%%%%%%%%%%%%%%%%%%%%%%%%%%%%%
\section{The instability}
%%%%%%%%%%%%%%%%%%%%%%%%%%%%%%%%%%%%%%%%%%%%%%%%%%%%%%%%%%%%%%%%% 

%%%%%%%%%%%%%%%%%%%%%%%%%%%%%%%%%%%%%%%%%%%%%%%%%%%%%%%%%%%%%%%%%%
\subsection{Time dependent perturbations}
%%%%%%%%%%%%%%%%%%%%%%%%%%%%%%%%%%%%%%%%%%%%%%%%%%%%%%%%%%%%%%%%% 

 We now investigate the stability of the RN solution, when embedded in the EYMCS model.
The functions $N$, $\sigma$, $w$ and $V$ satisfy a complicated set of  partial
differential equations which are found after replacing the ansatz (\ref{metric-gen}), (\ref{ansatz-gf})
in the field equations (\ref{field-eqs}).
However, for the purpose  of this paper, it is sufficient to study  
the equation for the magnetic gauge potential only~:
 \begin{eqnarray}
\label{m2}
%&&
%\frac{\partial }{\partial r}\left(\frac{r^3 }{\sigma}\frac{\partial V}{\partial r}+24 \kappa (w-\frac{1}{3}w^3 \right)=0,
\frac{\partial}{\partial r}\left(r N \sigma \frac{\partial w}{\partial r}\right)-
\frac{\partial}{\partial t}\left(\frac{r}{N} \frac{\partial w}{\partial t}\right)
+\frac{2\sigma w (1-w^2)}{r}+8 \kappa (1-w^2)\frac{\partial V}{\partial r}=0.
\end{eqnarray} 
Also, in what follows, we use the scaling symmetries of the
considered  
EYMCS system \cite{Brihaye:2010wp},
\cite{Brihaye:2011nr} to set $\alpha=1$.

The instability is found for values of the nA magnetic gauge potential close to the vacuum  value everywhere.
To consider fluctuations around the RN black hole, we only need to keep linear terms in 
$\delta w \equiv w+1$ (similar results are found for $\delta w \equiv w-1$), $\delta V\equiv V-V^{(RN)}$ and 
$\delta N\equiv N-N^{(RN)}$, $\delta \sigma\equiv \sigma-1$.
It turns out that the coupled equations separate and the equations for the metric components contains neither 
$\delta w$ nor  $\delta V$. The instablity occurs if there are solutions of the form
$\delta w=W(r) e^{\Omega t}$, with real $\Omega$.
From (\ref{m2}), one can see that $W(r)$ is a solution of the equation
\begin{eqnarray}
\label{eq-W}
\frac{d^2 W}{d\rho^2 }+4 (4 \kappa \frac{dV}{d\rho}-N)W-\Omega^2 W=0,
\end{eqnarray}

%%%%%%%%%%%%%%%%%%%%%%%%%%%%%%%%%%%%%%%%%%%%%%%%%%%%%%%%%%%%%%%%%%%%%%%%%% 
\setlength{\unitlength}{1cm}
\begin{picture}(8,6)
\put(-0.5,0.0){\epsfig{file=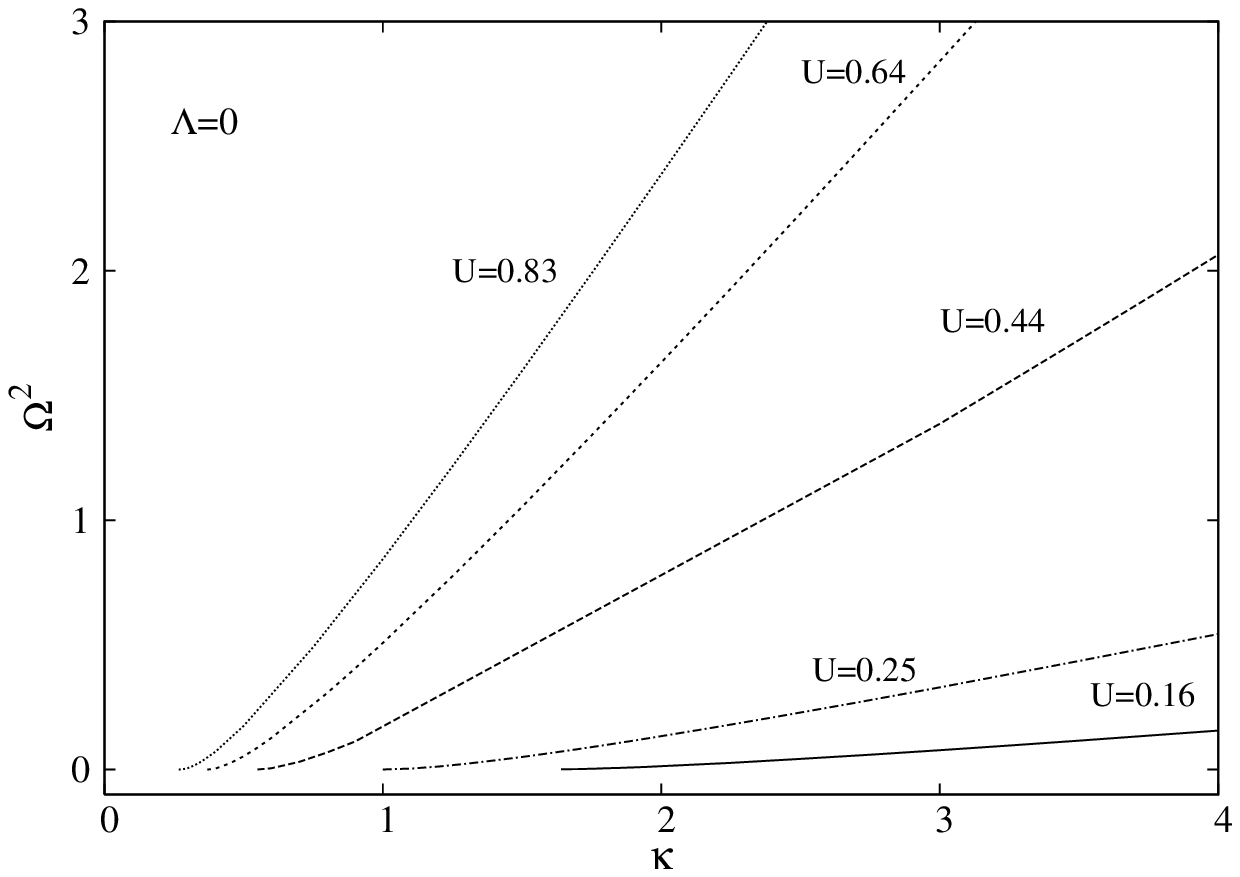,width=8cm}}
\put(8,0.0){\epsfig{file=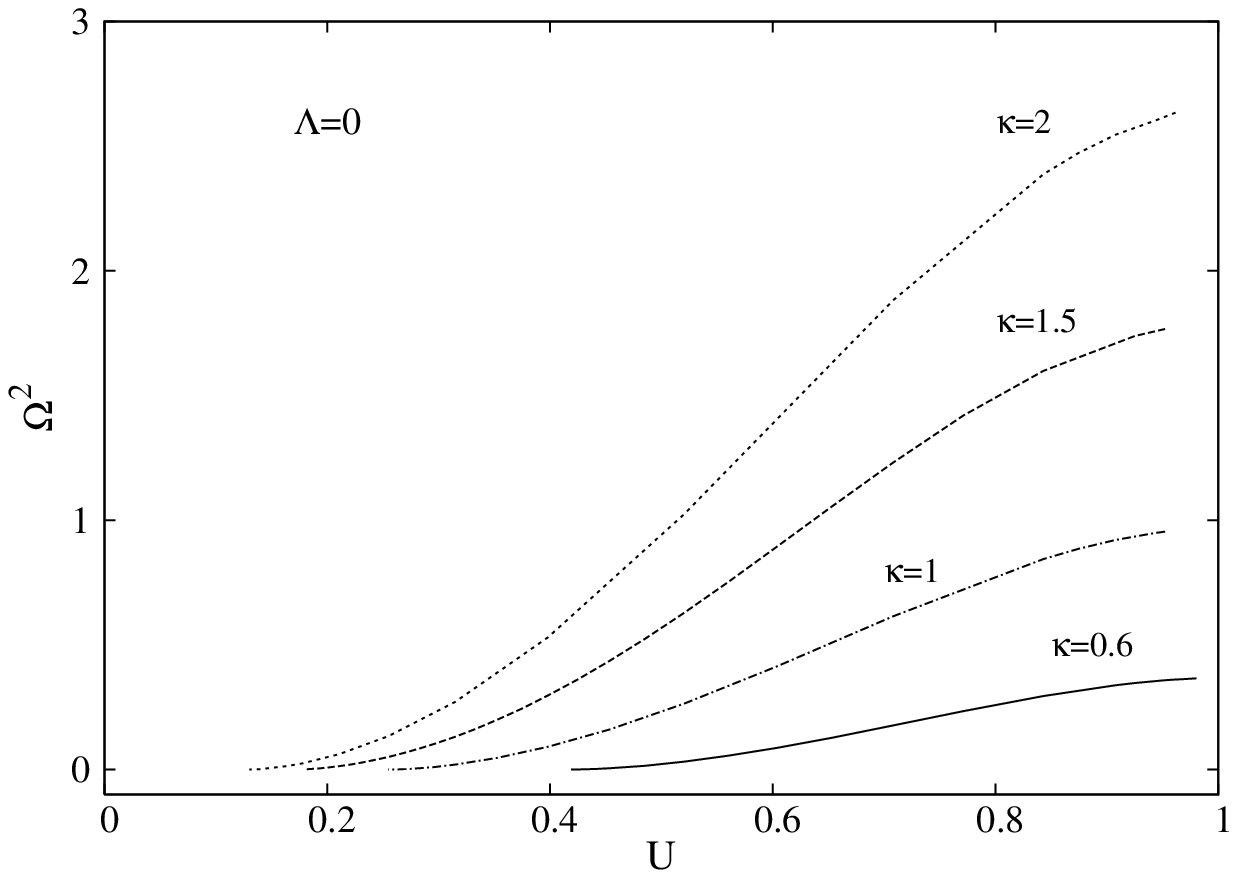,width=8cm}}
\end{picture}
\\
\\
{\small {\bf Figure 1.} The square of the frequency $\Omega^2$ is plotted as a function of the Chern-Simons coupling constant $\kappa$
  (left) and of the dimensionless parameter $U=Q/r_h^2$ (right).
   }
\vspace{0.5cm}
%%%%%%%%%%%%%%%%%%%%%%%%%%%%%%%%%%%%%%%%%%%%%%%%%%%%%%%%%%%%%%%%%%%%%%%%%%% 

%%%%%%%%%%%%%%%%%%%%%%%%%%%%%%%%%%%%%%%%%%%%%%%%%%%%%%%%%%%%%%%%%%%%%%%%%%% 
\setlength{\unitlength}{1cm}
\begin{picture}(8,6)
\put(-0.5,0.0){\epsfig{file=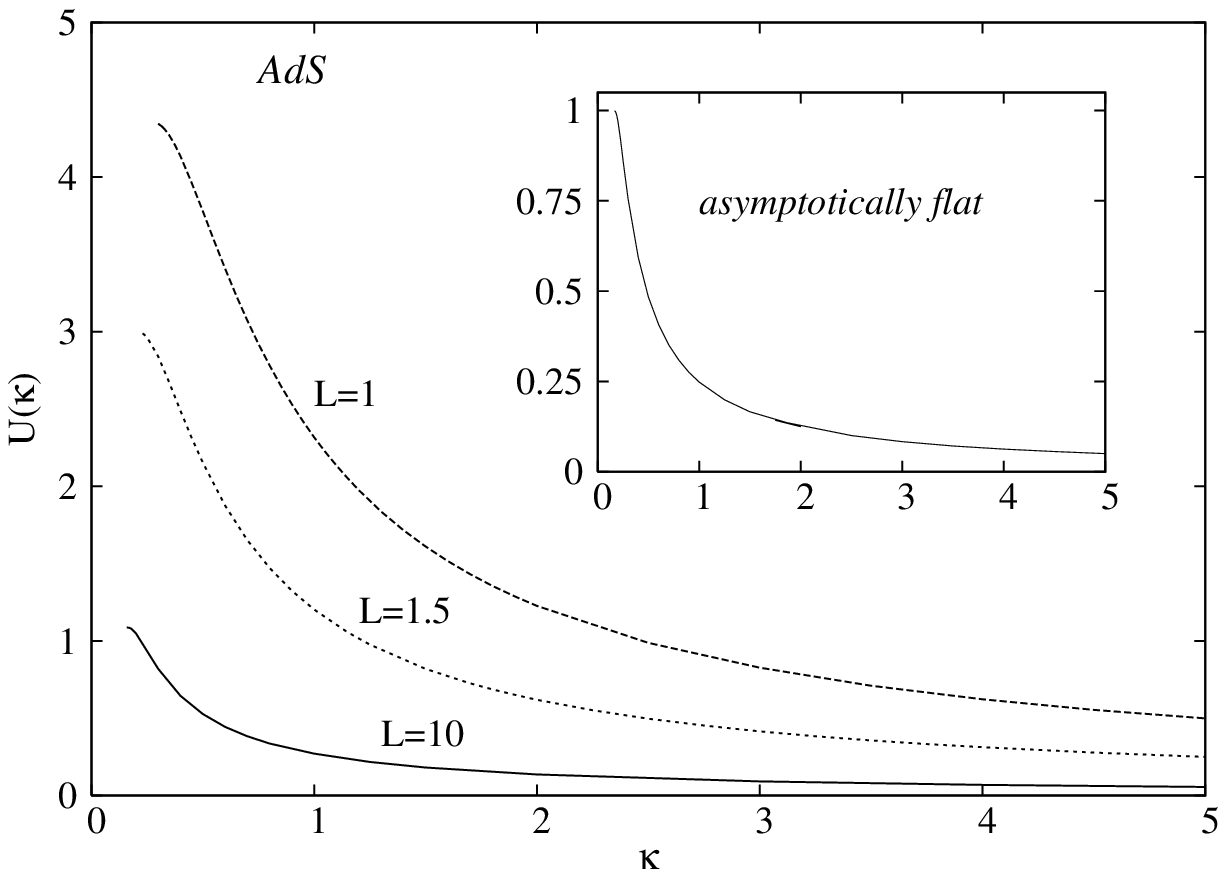,width=8cm}}
\put(8,0.0){\epsfig{file=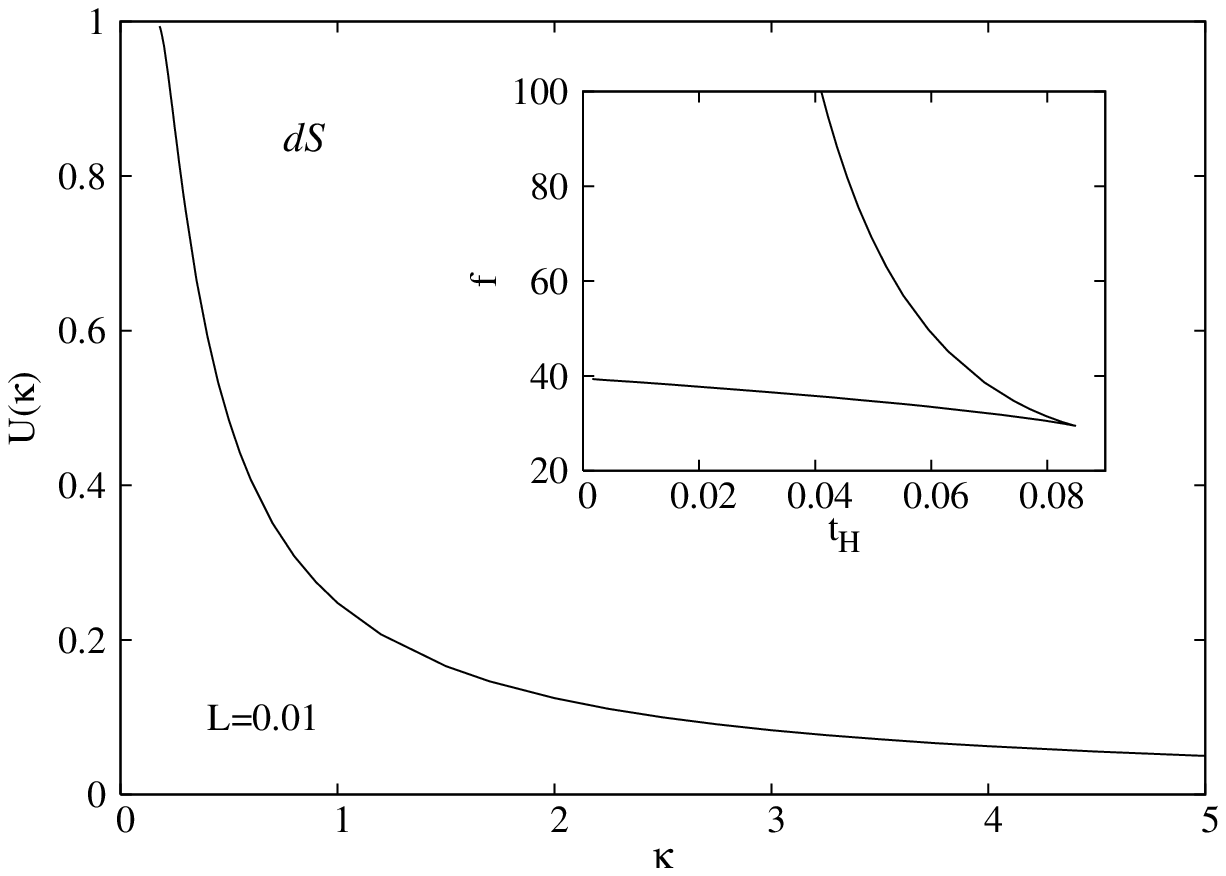,width=8cm}}
\end{picture}
\\
\\
{\small {\bf Figure 2.} The 'charge to radius' ratio $U(\kappa)=Q/r_h^2$ which 
characterizes the unstable Reissner-Nordstr\"om solution where a branch of non-Abelian
configurations emerges is shown as a function of 
the CS coupling constant $\kappa$ for solutions with a cosmological constant
$\Lambda \leq 0$ (left) and $\Lambda>0$ (right).
For de Sitter solutions, we show also the reduced  
free energy $f=({\cal M}-T_HS)/Q$ of the critical Reissner-Nordstr\"om configurations
as a function of reduced temperature $t_H=T_H\sqrt{Q}$, the CS coupling constant
there being the intrinsic parameter along that curve.
 }
\vspace{0.5cm}
%%%%%%%%%%%%%%%%%%%%%%%%%%%%%%%%%%%%%%%%%%%%%%%%%%%%%%%%%%%%%%%%%%%%%%%%%%%
\\
where we have introduced a new 'tortoise' coordinate  $\rho$ defined by $\frac{d r}{d\rho }=rN$,
such that the horizon appears at
$\rho  \to  -\infty$.
Close to the horizon, the physical solutions of (\ref{eq-W}) behave as 
\begin{eqnarray}
\label{sol1}
W(r)=w_0 e^{\Omega \rho} ,
\end{eqnarray}
with $w_0$ a free parameter. 
 As $r\to \infty$, the  approximate form of the solution is  
  \begin{eqnarray}
\label{sol2}
w(r) =e^{-\Omega r}\sqrt{r} \sum_{k=1}^\infty \frac{w_k}{r^k} ,
\end{eqnarray} 
for $\Lambda=0$ (where all coefficients $w_k$ are expressed in terms of $w_1$, $e.g.$ $w_2=w_1(15+8 M_0\Omega^2)/(8 \Omega)$), and 
  \begin{eqnarray}
\label{sol3}
w(r) =\sum_{k=2}^\infty \frac{w_k}{r^k},
\end{eqnarray} 
for $\Lambda\neq 0$ (again in terms of a single free coefficient $w_2$).

The perturbation $W(r)$ is determined by solving numerically the equation (\ref{eq-W})
with the boundary conditions (\ref{sol1})-(\ref{sol3}),
  by using a standard Runge-Kutta
ordinary  differential equation solver.
In practice we use  a shooting aproach in terms of the parameter $\Omega^2$.

 In Figure 1 we plot the frequency as a function of the CS coupling constant $\kappa$ for several values of $U=Q/r_h^2$
(left) and as a function of $U$ for several values of $\kappa$ (right).
One notices that, given $Q,r_h$, the RN black hole becomes unstable for all values of $\kappa>\kappa_{min}$,
with $\Omega\to 0$ as the minimal value of $\kappa$ is approached.
Equivalently, for fixed $\kappa$, a classical instability develops
if the horizon is sufficiently small ($i.e.$ if $U=Q/r_h^2$ is large enough).
Similar results were found in the presence of a cosmological term.

%%%%%%%%%%%%%%%%%%%%%%%%%%%%%%%%%%%%%%%%%%%%%%%%%%%%%%%%%%%%%%%%%%
\subsection{The marginally stable mode}
%%%%%%%%%%%%%%%%%%%%%%%%%%%%%%%%%%%%%%%%%%%%%%%%%%%%%%%%%%%%%%%%% 
For  $\Omega=0$, Eq. (\ref{eq-W}) takes the simple form
\begin{eqnarray}
\label{stab1}
 r(r N W')'-4(1-\frac{8\kappa Q}{r^2})W=0~.
\end{eqnarray}
The solutions of  (\ref{stab1}) 
are of particular interest since they indicate
the emergence of a branch of static, hairy black holes with a nonvanishing
magnetic field outside the horizon, $w(r) \neq \pm 1$. 
 The existence of these configurations can be understood as follows  \cite{Brihaye:2010wp},
\cite{Brihaye:2011nr}: 
the second term in (\ref{eq-W}) shows the presence of an effective mass term $\mu^2$ for $W$
near the horizon, with $\mu^2\sim 1-8 U(\kappa)$. 
All physical solutions have $\mu^2<0$,  with $Q/r_h^2=U(\kappa)$ being a monotonic
function of the CS coupling constant $\kappa$.

Although the equation (\ref{stab1}) does not appear to be solvable in terms
of known functions, one can construct an approximate solution
near the horizon and at infinity. 
As $r\to r_h$, the function $W(r)$ behaves as 
\begin{eqnarray}
\label{as1}
W(r)=b+\frac{2b r_h(r_h^2-8\kappa Q)L^2}{2\varepsilon r_h^6+(r_h^4-Q^2)L^2)}(r-r_h)+O(r-r_h)^2.
\end{eqnarray} 
For large $r$,   the approximate form of $W(r)$ is
\begin{eqnarray}
\label{as2}
 W(r)=\frac{J}{r^2}+\frac{J}{3\varepsilon}(M_0-4\kappa Q)\frac{L^2}{r^6}+\dots,~~ {\rm for}~~\Lambda \neq 0 ,
\end{eqnarray}
and
\begin{eqnarray}
\label{as3} 
 W(r)=\frac{J}{r^2}+\frac{2J}{3 }(M_0-4\kappa Q)\frac{1}{r^4}+\dots,~~ {\rm for}~~\Lambda =0 .
\end{eqnarray}
In the above relations, $b$ and $J$ are free parameters (in practice one sets $b=1$ without any loss of
generality).
 
 Solutions interpolating between the asymptotics (\ref{as1}) and  (\ref{as2}), (\ref{as3})
are constructed numerically\footnote{In this work we restrict our 
study to solutions with a monotonic behaviour of $W(r)$.}.
In practice, the input parameters are $\kappa$ and $Y=r_h^2/L^2$.
The "charge to radius" ratio $U=Q/r_h^2$ results from the numerical output. 
In Figure 2 we show the profile of $U(\kappa)$ for several values of the cosmological
length scale $L$ (the event horizon radius is taken $r_h=1$).
    
An analytic expression of $U(\kappa)$ can be found by  
matching at some intermediate 
point\footnote{In this approach, a new radial coordinate is introduced, $x=r_h/(1-x)$, with $0\leq x\leq 1$.}
 the expansion of $W(r)$ (and its first derivative) at the horizon, (\ref{as1}),
to that at infinity, (\ref{as2}), (\ref{as3}).  
Surprisingly, it turns out that the resulting expression
\begin{eqnarray}
\label{approx}
U=\frac{Q}{r_h^2}=-4 \kappa+\sqrt{2+16\kappa^2+2\varepsilon Y},
\end{eqnarray}  
provides a good approximation of the numerical data (typically with several percent error) and 
gives also an analytic explanation for some numerical results found in \cite{Brihaye:2010wp} for $\Lambda=0$.
For example, for large $\kappa$ one finds
$U\simeq (1+\varepsilon Y)/(4\kappa)$, with $U\kappa=1/4$ for $\Lambda=0$.
Also, the condition for tachionic mass gives a minimal value of the coupling constant $\kappa$
\begin{eqnarray}
\label{kmin}
\kappa>\kappa_{min}=\frac{1}{8\sqrt{1+2\varepsilon Y}},
\end{eqnarray} 
which implies $\kappa \geq 1/8$ for $\Lambda=0$.

The knowledge of the function $U$ allows us to reconstruct, after replacing in (\ref{funct-U1}),
the thermodynamical parameters of the unstable RN solution as a function of $\kappa,Y$. 
For example, when embedded in EYMCS theory with $\Lambda=0$, the RN black holes 
possess a marginaly stable mode for a value of the dimensionless ratio
\begin{eqnarray}
\label{rel1}
T_H^3 A_H=\frac{1}{4\pi}
\left(
1-(4\kappa -\sqrt{2+16\kappa^2})^2
\right)^2
,
\end{eqnarray} 
 while $t_H=T_H\sqrt{Q} \to 0$ as $\kappa \to 1/8$.

%%%%%%%%%%%%%%%%%%%%%%%%%%%%%%%%%%%%%%%%%%%%%%%%%%%%%%%%%%%%%%%%%%
\section{Black holes with non-Abelian hair}
%%%%%%%%%%%%%%%%%%%%%%%%%%%%%%%%%%%%%%%%%%%%%%%%%%%%%%%%%%%%%%%%% 
 
The instability of the RN solution pointed out in the previous section can be viewed as an indication 
of the existence of a branch of non-Abelian solutions with nontrivial
magnetic non-Abelian fields outside the horizon.
These configurations appear to be the natural end points to
which the instability of the RN solution leads.

Such solutions  are constructed  nonperturbatively by solving numerically 
the set of equations   following from (\ref{field-eqs})~:
\begin{eqnarray}
\label{eqs}
&&m'=\frac{1}{2} 
\left(
3r \bigg(
N w'^2 
+\frac{(w^2 -1)^2}{r^2}
\bigg)
+\frac{r^3}{\sigma^2}  V'^2 
\right),
~~\frac{\sigma'}{\sigma}=\frac{3 w'^2 }{2r},
\\
\nonumber
&&(r\sigma Nw')'=
\frac{2 \sigma  w(w^2 -1)}{r }
+8\kappa 
 V'(w^2 -1) ,
~~~\big (\frac{r^3 V'}{\sigma}\big )'= 24 \kappa (w^2 -1)w',
\end{eqnarray}
 where we set $\alpha^2 \equiv 16\pi G/(3e^2)=1$ and define $m(r)$ according to
\begin{eqnarray}
\label{N}
N(r) = 1-\frac{m(r)}{r^2}+\varepsilon \frac{r^2}{L^2}.
 \end{eqnarray}
  The function $m(r)$ is (up to a constant factor) the local mass-energy density.
 
  With the scale fixing adopted, the reduced equations depend on
 two input parameters -- the (A)dS length
scale $L$ and the CS coupling constant $\kappa$.
The invariance of (\ref{eqs}) to changing the signs of $w,V$ and $\kappa$
allows to study solutions with strictly positive values of $\kappa$ only.

We notice that the equation for $V(r)$ has the first integral \cite{Brihaye:2011nr}
\begin{eqnarray}
\label{1SU2U1}
V'=\frac{2\sigma}{r^3}\big(Q+4 \kappa (w-2)(w+1)^2\big), 
\end{eqnarray}
with $Q$ an integration constant which is taken to be positive without any loss of generality.

For any sign of $\Lambda$, the expression of the solutions as $r\to \infty$ as imposed by the
finite mass/energy requirement, is\footnote{Choosing $w(\infty)=1$ leads to similar results.}
\begin{eqnarray}
\label{exp-inf}
 m(r)=M-\frac{Q^2+3\varepsilon J^2/L^2}{r^2} +\dots,
~~
\sigma(r)=1-\frac{ J^2}{r^6}+\dots,
~~ w(r)=- 1+\frac{ J}{r^2}+\dots,~~V(r)=V_0-\frac{Q}{r^2}+\dots~,~~
\end{eqnarray} 
where $J,~M,~V_0$ are parameters given by numerics which fix all higher order terms.
 
We want the line element (\ref{metric-gen}) to describe a black hole, with an event horizon located at $r=r_h>0$. 
In the vicinity of the event horizon, one can write a formal power-series expansion  of the solution  
\begin{eqnarray}
\nonumber
&&m(r)=m_0+\sum_{k=1}^{\infty}m_k (r-r_h)^k,~~\sigma(r)=\sigma_h+\sum_{k=1}^{\infty}\sigma_k (r-r_h)^k,~~
%&&m(r)=m_0+m_1(r-r_h)+\dots,
%~
%\sigma(r)=\sigma_h+\frac{3 \sigma_h w_1^2}{2 r_h}(r-r_h)+\dots,
\\
\label{exp-eh}
&&w(r)=w_h+\sum_{k=1}^{\infty}w_k (r-r_h)^k,~~
 V(r)=\sum_{k=1}^{\infty}v_k (r-r_h)^k.
\end{eqnarray} 
All coefficients in the above series are determined through reccurence relations by $w_h$ and $\sigma_h$.
 One finds $e.g.$
\begin{eqnarray}
\nonumber
&&
v_1=\frac{2\sigma_h}{r_h^3}(Q+4\kappa (1+w_h)^2(w_h-2)),
\\
&&
m_1= \frac{1}{2r_h^3}(4Q^2+32 \kappa Q(1+w_h)^2(w_h-2)+
64 \kappa^2(w_h-2)^2(1+w_h)^4+3r_h^2(w_h^2-1)^2),
\\
\nonumber
&&
w_1=  \frac{2L^2(4 \kappa r_h v_1 +\sigma_h w_h)(w_h^2-1)}{(4\varepsilon r_h^3-(m_1-2r_h)L^2)\sigma_h}.
\end{eqnarray} 
Thus this expansion (assuming that its radius of convergence is nonzero) defines a two-parameter family of
local solutions labeled by the values of $w(r)$ and $\sigma(r)$  on the horizon.

The only conserved quantities associated with these solutions are the mass ${\cal M}$ and the electric charge ${\cal Q}$
\begin{eqnarray} 
\label{def-M}
{\cal M}=\frac{3\pi}{8 G}M,~~{\cal Q}=\frac{4 \pi^2 }{e}Q.
\end{eqnarray} 
Other quantities of interest are the chemical potential $\Phi=\frac{V_0}{e}$, the Hawking temperature $T_H$,
and the entropy $S$
\begin{eqnarray} 
T_H= \frac{1}{4 \pi} \sigma(r_h) N'(r_h),~~S=\frac{A_H}{4 G}=\frac{\pi^2 r_h^3}{2G}.
\end{eqnarray} 
One should note that, similar to other nA hairy black holes, 
there is no conserved quantity associated with the parameter 
$J$ which appears in the large-$r$ asymptotics (\ref{exp-inf}).

The dS solutions have some special features due to the presence of a cosmological horizon at $r=r_c>r_h$.
(See the discussion in Section 4.)

%%%%%%%%%%%%%%%%%%%%%%%%%%%%%%%%%%%%%%%%%%%%%%%%%%%%%%%%%%%%%%%%%%
\subsection{The cases $\Lambda \leq 0$}
%%%%%%%%%%%%%%%%%%%%%%%%%%%%%%%%%%%%%%%%%%%%%%%%%%%%%%%%%%%%%%%%% 

The asymptotically flat solutions of this model have been discussed at length in \cite{Brihaye:2011nr}.
The results there indicated that the nA solutions emerge as perturbations of the RN black holes
for a critical charge to radius ratio $U$, which is fixed by the CS coupling constant $\kappa$.
%Also, the properties of the solutions depend  on the value of the CS coupling constant $\kappa$.
Hairy black holes appear to exist for any value of $\kappa \geq 1/8$.
For a fixed $r_h$, the value of  the electric charge parameter $Q$ is also important,
some basic features of the solutions depending on whether $Q$ is less or greater than  
\begin{eqnarray}
\label{Qc}
Q^{(c)}=16\kappa.
 \end{eqnarray} 
For fixed $\kappa$, a plot of the reduced horizon area $a_H$
$vs.$ the reduced Hawking temperature $t_H$ reveals the existence of several branches of solutions.
For $Q\neq Q^{(c)}$, the picture is rather similar to that valid for RN black holes \cite{Brihaye:2011nr}.
The temperature reaches its maximum at some intermediate value of the event horizon radius,
an extremal black hole being approached for a minimal value of $r_h>0$.
The lower branch of solutions has positive specific heat. Solutions with  $Q=Q^{(c)}$ are special, since one 
finds a single branch of thermally unstable configurations.
These solutions emerge again from a critical RN solution with  $r_h=\sqrt{Q^{(c)}/U(\kappa)}$
and can be continued for an arbitrarily small value of the event horizon radius.
As $r_h\to 0$, these  black holes
%%%%%%%%%%%%%%%%%%%%%%%%%%%%%%%%%%%%%%%%%%%%%%%%%%%%%%%%%%%%%%%%%%%%%%%%%% 
\setlength{\unitlength}{1cm}
\begin{picture}(8,6)
\put(-0.5,0.0){\epsfig{file=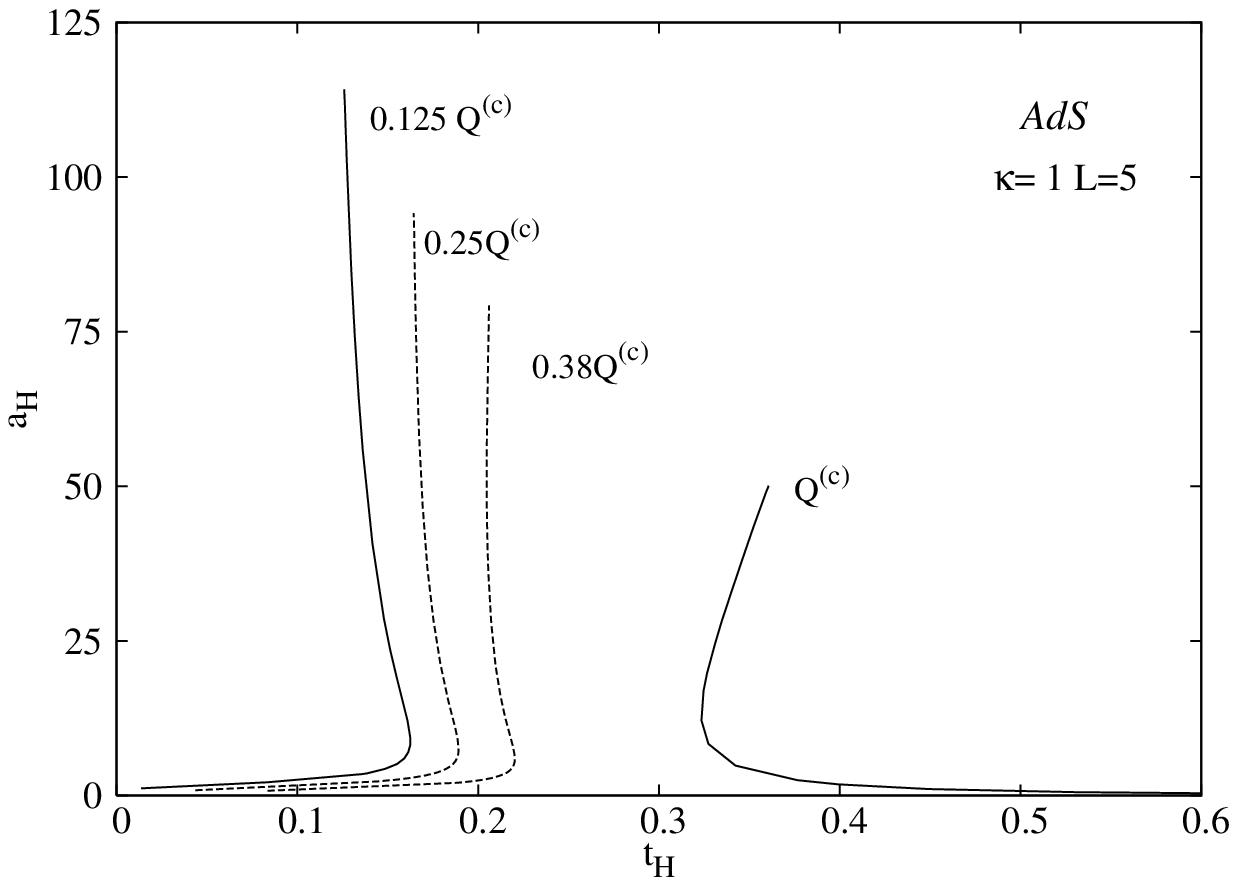,width=8cm}}
\put(8,0.0){\epsfig{file=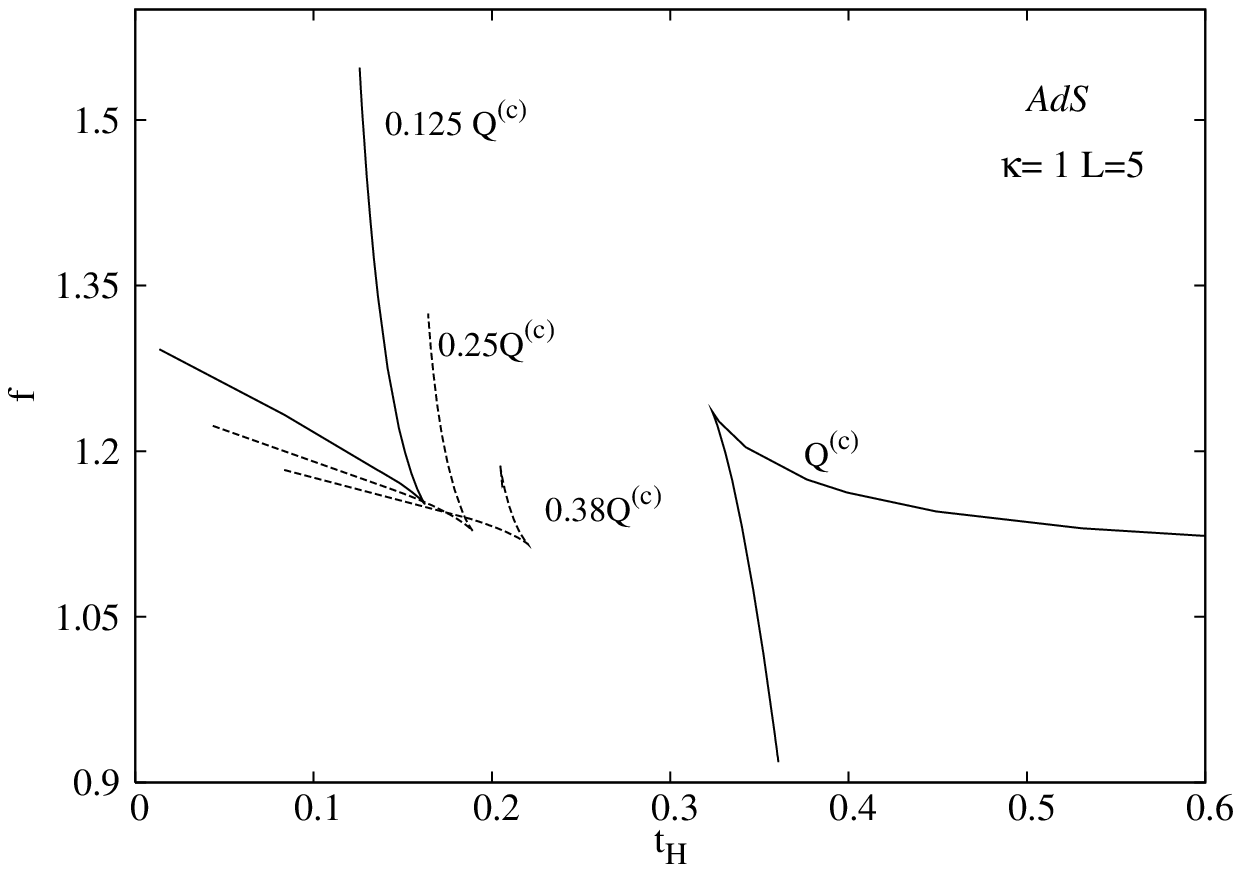,width=8cm}}
\end{picture}
\\
\\
{\small {\bf Figure 3.} The reduced event horizon area $a_H=A_H/Q^{3/2}$ (left) and 
free energy $f=({\cal M}-T_H S)/Q$ (right) are plotted $vs.$ the reduced temperature $t_H=T_HQ^{1/2}$ for
AdS branches of non-Abelian  solutions with several values of the electric
charge and a given value of $\kappa$. 
   }
\vspace{0.5cm}
%%%%%%%%%%%%%%%%%%%%%%%%%%%%%%%%%%%%%%%%%%%%%%%%%%%%%%%%%%%%%%%%%%%%%%%%%%% 

%%%%%%%%%%%%%%%%%%%%%%%%%%%%%%%%%%%%%%%%%%%%%%%%%%%%%%%%%%%
 \setlength{\unitlength}{1cm}
\begin{picture}(8,6)
\put(3,0.0){\epsfig{file=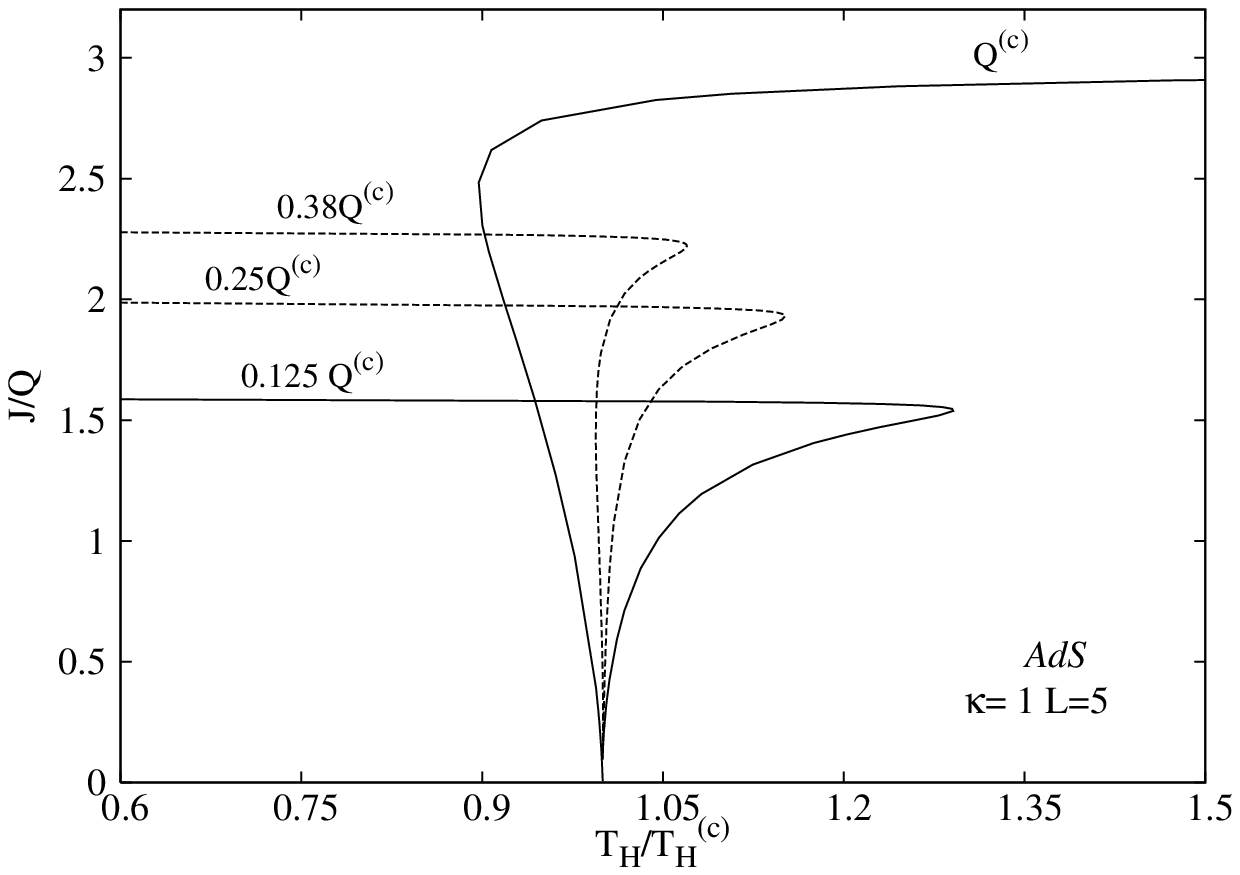,width=8cm}} 
\end{picture}
\\
\\
{\small {\bf Figure 4.}
The  order parameter $J$ which enters the asymptotics
of the magnetic gauge potential at infinity 
is shown as a function of the Hawking temperature $T_H$
for a fixed value of the Chern-Simons coupling constant
$\kappa$ and several values of the electric charge.
Note that $T_H$ is normalized with respect the temperature of the critical
Abelian solution where a branch of non-Abelian
solutions emerges as a perturbation. }
\vspace{0.5cm}
%%%%%%%%%%%%%%%%%%%%%%%%%%%%%%%%%%%%%%%%%%%%%%%%%%%%%%%%%%%
\\
 approach a set of globally regular particle-like solutions.
More details on the properties of the $\Lambda=0$ configurations togheter with typical profiles can be
found in \cite{Brihaye:2010wp}, \cite{Brihaye:2011nr}. 
 
The basic features of the AdS solutions were discussed in \cite{Brihaye:2009cc}.
However, the main emphasis there was on the solutions of the full $SO(6)$
model, featuring two extra gauge potentials apart from $w,V$.
The $SU(2)\times U(1)$ solutions of interest here corresponds to  the family
of solutions which was referred in \cite{Brihaye:2009cc} as 
the 'fundamental branch of EYMCS solutions'. 
We have reexamined the properties and the basic features of these solutions and 
have found that they emerge indeed as perturbations of the RNAdS configurations.
The solutions exist for values of $\kappa$ above some minimal value which is well approximated by (\ref{kmin}).  

%%%Different from the $\Lambda=0$ case, 
 We did not investigate in a systematic way
the thermodynamics of the solutions corresponding to $\Lambda \neq 0$.
(Recall that the presence of a negative cosmological constant leads to a complicated 
thermodynamical picture already for RN black holes \cite{Chamblin:1999tk}, \cite{Chamblin:1999hg}).
The set of solutions  investigated so far have a relatively small ratio $Y=r_h^2/L^2$
and  reveal a picture which is rather
similar to the asymptotically flat case.
For $Q\neq Q^{(c)}$, a temperature-area plot indicates the existence of two branches of solutions (see Figure 3
for typical results).
The upper branch of solutions emerges from a critical RN black hole with  $r_h=\sqrt{Q/U(\kappa)}$
and is thermally unstable, $i.e.$ the temperature increases for decreasing $r_h$.
The lower branch always possesses a positive specific heat, the Hawking temperature vanishing there 
for a minimal value  $r_h^{(min)}$  of the event horizon radius. As $r\to r_h^{(min)}$,
a zero temperature non-Abelian black hole solution with a regular horizon is approached\footnote{  
The existence of these extremal configurations is supported by an exact $AdS_2\times S^3$
solution of the Eqs. (\ref{eqs}) \cite{Brihaye:2009cc}.
This exact solution describes the near horizon geometry of an extremal black hole with nA hair.}.
As noticed already in \cite{Brihaye:2009cc}, the solutions with $Q=Q^{(c)}=16 \kappa$
exist for arbitrarily small values of $r_h$, a particle-like solution with a regular origin 
being approached as  $r_h\to 0$.

In Figure 4 we plot the dimensionless quantity $J/Q$ (with $J$ 
the order parameter which enters the first relevant term in the large$-r$ expansion
of the magnetic gauge potential) as a function of the reduced   Hawking temperature  
and several values of the electric charge parameter.  One can see that for $Q<Q^{(c)}$, $J$
approaches a constant, nonvanishing value for small enough values of $T_H$. 

In the discussion of both asymptotically flat and AdS solutions, we have restricted
to configurations with a monotonic behaviour of the magnetic gauge potential $w(r)$.
Solutions with  local extrema of $w(r)$ do exist, but it is likely that they are always thermodynamically disfavoured 
because spatial oscillations in $w$ increase the total mass. 
Also, as expected (since they start as perturbationS around RN black holes), for both Minkowski and AdS asymptotics, 
we have found nA solutions where $w(r)$ has no nodes.

%%%%%%%%%%%%%%%%%%%%%%%%%%%%%%%%%%%%%%%%%%%%%%%%%%%%%%%%%%%%%%%%%%%%%%
\subsection{$\Lambda>0$: de Sitter solutions}
%%%%%%%%%%%%%%%%%%%%%%%%%%%%%%%%%%%%%%%%%%%%%%%%%%%%%%%%%%%%%%%%%%%%%%%

Similar solutions should exist for positive values of $\Lambda$, 
as indicated by the presence of an unstable mode also in that case.
However, the situation is much more involved in this case due to the presence of 
a cosmological horizon. 

The dS configurations are found again by solving numerically the Eqs. (\ref{eqs}). 
Some of the basic features of these solutions are generic for any black hole in a dS background.
For $\Lambda>0$, the solutions possess a cosmological horizon located at $r = r_c > r_h$.  
In this case, one needs another set of boundary conditions at the cosmological horizon,
with $N(r_c)=0$, $N'(r_c)<0$, $\sigma(r_c)>0$.
A formal power series similar to (\ref{exp-eh}) can also be written at $r=r_c$,
the only free parameters being $w(r_c)$ and $\sigma(r_c)$.
Moreover, both the event and the cosmological horizons have their own surface gravity (and hence different
temperatures.) Outside the cosmological horizon $r$ and $t$ changes character
($i.e.$ $r$ becomes a timelike coordinate for $r > r_c$). 
A nonsingular extension across this null surface can be found just as at the event horizon of a black hole. 
The regularity assumption implies that all curvature invariants at $r = r_c$ are finite. Also, all matter
functions and their first derivatives extend smoothly through the cosmological horizon, $e.g.$ in a similar way
as the $U(1)$ electric potential of a RNdS solution.

The mass and electric charge of the solutions are computed outside the cosmological horizon, as $r\to \infty$
(note that the asymptotic expansion (\ref{exp-inf}) still holds for $\Lambda>0$, although
$N(r)<0$ for $r>r_c$). Similar $e.g.$ to the Schwarzschild or RN black holes, the  total
mass, as found by employing the quasilocal formalism in \cite{Ghezelbash:2001vs}, is ${\cal M}=-\frac{3\pi}{8 G}M$ 
(thus with a minus sign as compared to the definition (\ref{def-M}),
which is well-known feature of the dS solutions).
 
The field equations were solved in this case by employing a collocation method for boundary-value ordinary
differential equations, equipped with an adaptive mesh selection procedure \cite{colsys}.
As in the Einstein-Maxwell case, the cosmological horizon radius $r_c$ is a function of $\Lambda$. 
In practice, we solved the equations first on the interval $[r_h, r_c]$, choosing $r_c$ by hand and imposing
regularity conditions of the solutions at $r = r_c$. This allows to determine the values of the functions
$(m, \sigma,w,w',V)$ at $r = r_c$, as well as the numerical value of $\Lambda$ corresponding to the choice of
cosmological horizon. The integration on $[r_c,\infty)$ can then be performed as a second step, leading to the
knowledge of the solution on the full $r-$interval.

Our numerical results indicate the existence of nontrivial solutions of the equations 
(\ref{eqs}) approaching as $r\to \infty$ the dS background. The profile of a typical black hole solution
with nA hair is shown in Figure 5. One notices that the Yang-Mills fields do not vanish at $r=r_c$
and persist  beyond the cosmological horizon. Thus
these solutions may be thought of as cosmic colored black holes.
Also, similar to the $\Lambda\leq 0$ case, solutions with a regular origin are found for
the value of the electric charge $Q=16\kappa$ only.

\newpage
 
%%%%%%%%%%%%%%%%%%%%%%%%%%%%%%%%%%%%%%%%%%%%%%%%%%%%%%%%%%%
 \setlength{\unitlength}{1cm}
\begin{picture}(8,6)
\put(3,0.0){\epsfig{file=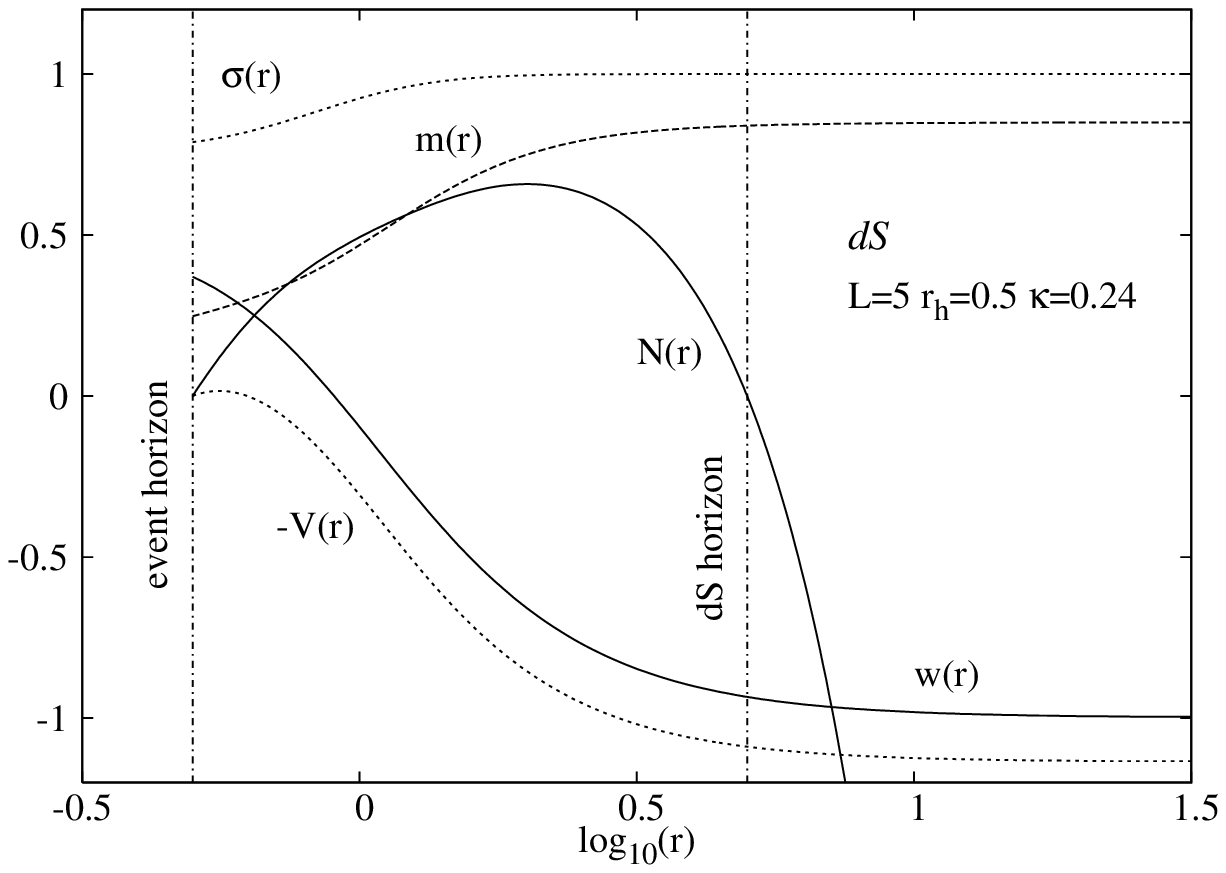,width=8cm}} 
\end{picture}
\\
\\
{\small {\bf Figure 5.} The profile  of a typical hairy black hole solution with de Sitter asymptotics is 
presented as a function of the radial coordinate $r$. $m(r)$ and $\sigma(r)$ are metric functions, while $V(r)$
and $w(r)$ are electric and magnetic gauge potentials, respectively.
}
\vspace{0.5cm}
%%%%%%%%%%%%%%%%%%%%%%%%%%%%%%%%%%%%%%%%%%%%%%%%%%%%%%%%%%%

%%%%%%%%%%%%%%%%%%%%%%%%%%%%%%%%%%%%%%%%%%%%%%%%%%%%%%%%%%%%%%%%%%%%%%%%%%% 
\setlength{\unitlength}{1cm}
\begin{picture}(8,6)
\put(-0.5,0.0){\epsfig{file=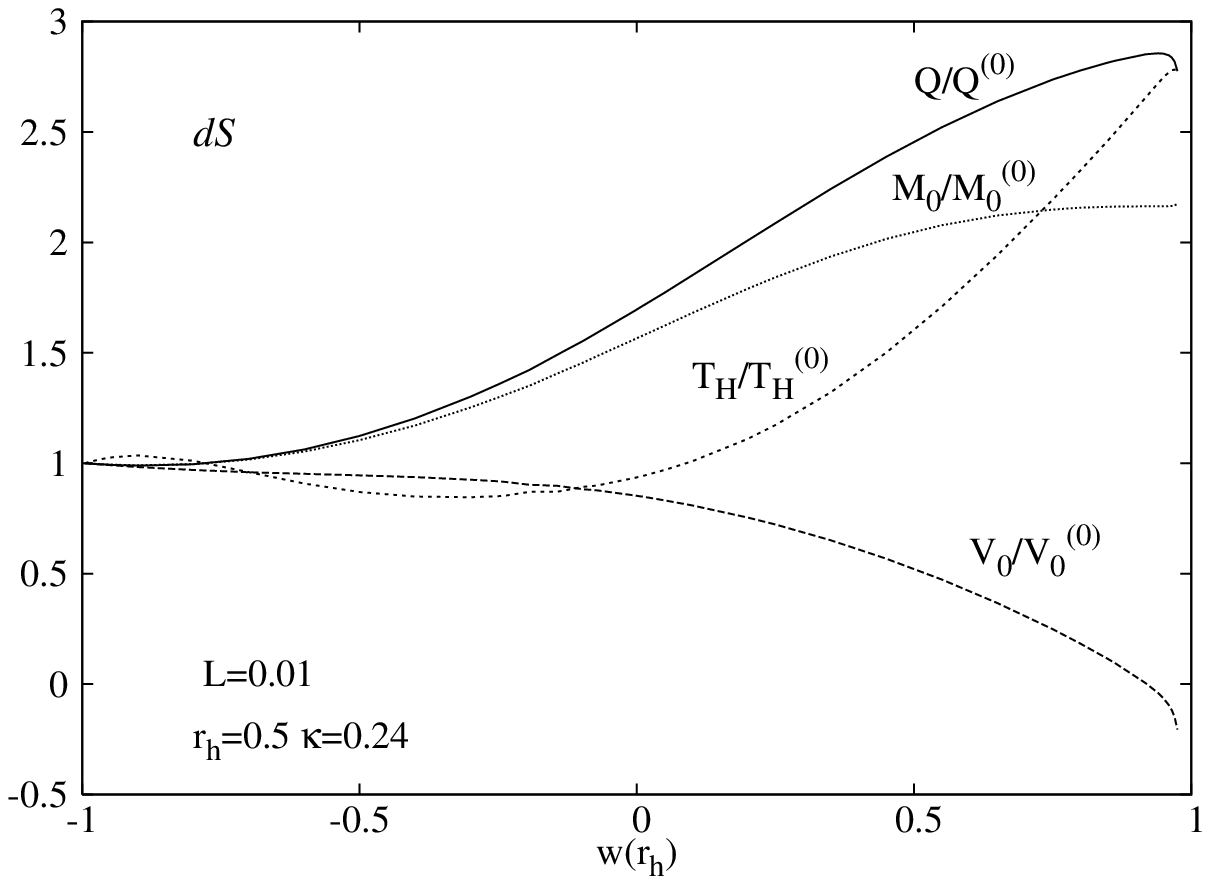,width=8cm}}
\put(8,0.0){\epsfig{file=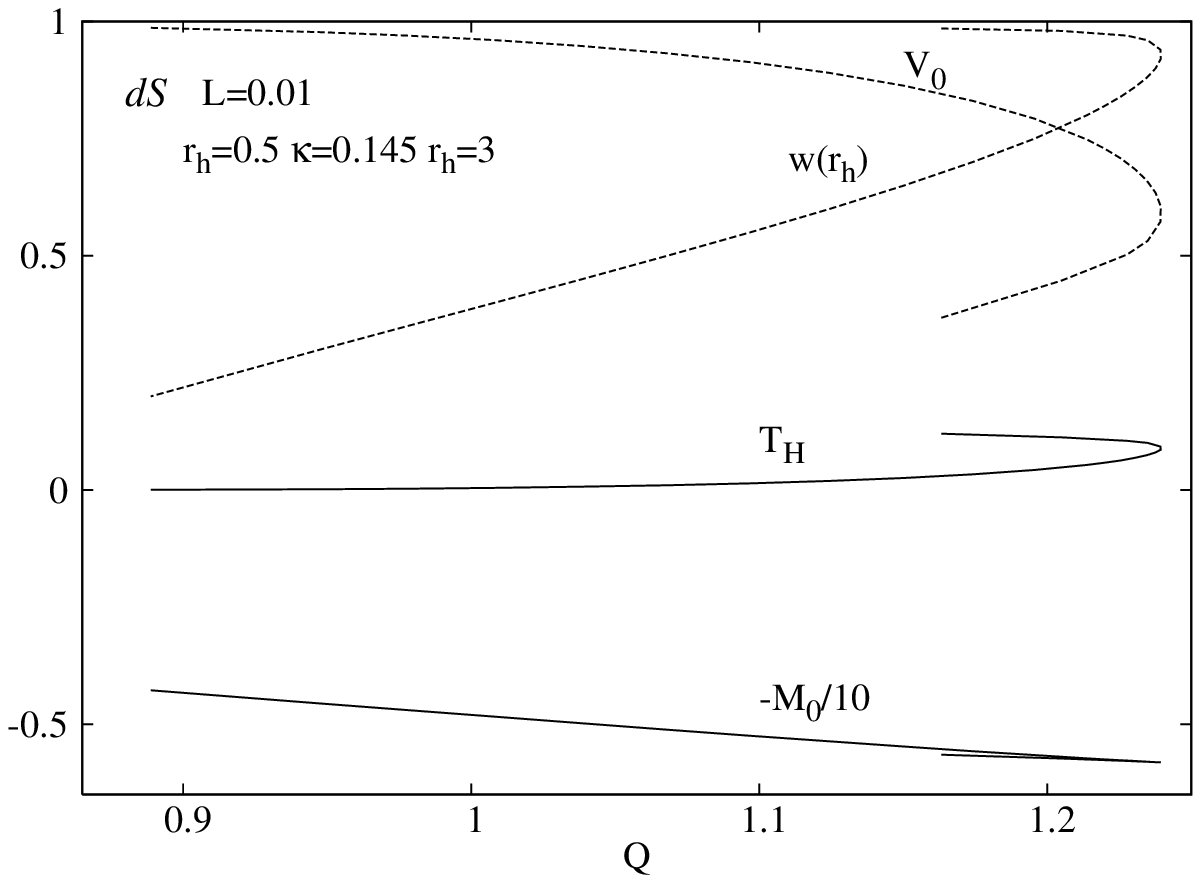,width=8cm}}
\end{picture}
\\
\\
{\small {\bf Figure 6.}
Several relevant parameters are plotted as a function of the magnetic gauge potential on the horizon (left) and of the 
electric charge parameter  $Q$ (right) for de Sitter black hole solutions with a fixed value of the Chern-Simons coupling
constant $\kappa$ and a given event horizon radius.   }
\vspace{0.5cm}
%%%%%%%%%%%%%%%%%%%%%%%%%%%%%%%%%%%%%%%%%%%%%%%%%%%%%%%%%%%%%%%%%%%%%%%%%%% 

However, determining the pattern of the dS solutions in terms of the input parameters $\kappa,L$ 
represents a very complex task which is outside the scope of this paper.
Instead, we have studied particular sets of solutions for several choices of $\kappa,L$ only.
Several parameters characterizing these solutions are presented on Figure 6.
There we show first the dependence of the solutions on the value of the
magnetic gauge potential on the horizon for fixed values of the CS coupling constant $\kappa$,
event horizon radius $r_h$ and cosmological length scale $L$.  
The quantities plotted there are normalized with respect to those of
the critical RN-dS solution which possess a marginally stable mode. The dependence of the solutions on
the value of the electric charge is also shown in Figure 6 (right) 
for a different set of relevant parameters.

When studying the pattern of solutions for a fixed value of the  electric charge $Q$ and  
increasing the event horizon radius $r_h$, we find that generically the branch of nA 
black holes ends up in
an extremal  solution  for $r_h \to r_h^{c}$. 
However, the  picture is quite complicated and seems to depend in a 
nontrivial way on the value of $\kappa$. For example, for some values of $\kappa$, 
we have found that
when $w_h$ becomes sufficiently small, the metric function $N(r)$ develops a local minimum 
at some intermediate value of the radial variable, say $r = r_0 > r_h$ and $r_0<r_c$.
The value $N(r_0)$ tends to zero for some minimal value $w_h = w_{h,c}$ and the solution therefore
bifurcates into an extremal RN black hole in the limit $w_h \to w_{h,c}$. 
Inspecting the matter fields, we further note that, for the limiting solution,
the gauge field function $w(r)$ is not constant
 for $r \in [r_h, r_0)$. However, the magnetic field vanishes for $r\in [r_0,\infty)$, $i.e.$ $w^2(r)=1$, and 
at the same time, the electric potential $V$  approaches the one corresponding to a RNdS black hole.
As a consequence, an extremal RNdS black hole is formed for $r\in [r_0,\infty]$. 
This kind of bifurcation resembles a 
phenomenon observed some years ago in the context of gravitating monopole and black holes \cite{Breitenlohner:1994di}.
The control parameter is there $\tilde \alpha = G/ v^2$ (with $G$ the Newton's constant, and $v$  the
expectation value of the  Higgs field).
It was shown that, when $\alpha$ becomes large enough, the metric function $g_{rr}(r)$
also develops a local minimum at some intermediate value $r=r_0$ and the gravitating monopole (or monopole
black hole) bifurcates into an extremal RNdS solution for $r \in [r_0,\infty]$.
Here the phenomenon occurs in the absence of a Higgs field, the matter field is purely Yang-Mills type
and the control parameter is $\kappa$, which like the Higgs VEV, is a dimensionful quantity.
 
We conjecture that, similar to the AdS case, a positive cosmological case allows for the
existence of a larger set of spherically symmetric solutions, with
four non-trivial gauge functions ($i.e.$ for the full group $SO(6)$).
As argued in \cite{Brihaye:2011nr}, these solutions are unlikely to exist in 
the asymptotically flat case.

We close the discussion here by noticing that for $\kappa=1/8$,
by using the generation technique proposed in \cite{Kastor:1992nn},
one finds the following exact solution  
\begin{eqnarray}
\label{exact-solution-dS}
ds^2=\frac{e^{2t/L}}{F(r,t)} (dr^2+r^2d\Omega_3^2)- F^2(r,t)dt^2,~
w(r)=\frac{J-2 r^2}{J+2 r^2},~
V(r,t)= {F(r,t)},
\end{eqnarray}
where $F(r,t)^{-1}= 1+ e^{-2t/L}\left(\frac{Q-2}{r^2}+\frac{2(r^2+J)}{(r^2+J/2)^2}\right)$ and $\Lambda=6/L^2$.
This solution is written in an inflationary coordinate system 
and describes a non-Abelian deformation of the $d=5$ extremal RN-de Sitter black holes discussed $e.g.$ in  
\cite{Astefanesei:2003gw}.
The time-dependence enters in a natural manner in this solution and for large $r$, the dS
metric in planar coordinates is approached.  
To some extent, the properties of the nA solution (\ref{exact-solution-dS}) can be discussed in a manner similar to
the case of RNdS black hole written in planar coordinates, see $e.g.$ \cite{Astefanesei:2003gw}.
In this approach, the mass is computed at far past or far future infinity (depending on the sign of $L$) 
by using  the quasilocal tensor of Brown and York \cite{Brown:1992br} (augmented by suitable
boundary counterterms \cite{Ghezelbash:2001vs}),
the resulting expression being ${\cal M}=-8\pi^2$.

%%%%%%%%%%%%%%%%%%%%%%%%%%%%%%%%%%%%%%%%%%%%%%%%%%%%%%%%%%%%%%%%%%%%%%
\section{Further remarks and conclusions}
%%%%%%%%%%%%%%%%%%%%%%%%%%%%%%%%%%%%%%%%%%%%%%%%%%%%%%%%%%%%%%%%%%%%%%%
In  this paper we have presented arguments that the 
$d = 5$ electrically charged  RN black holes
become unstable when $SU(2)$ nA fields are introduced in the action, with a specific coupling to the $U(1)$ field.
This instability allows for dressing RN solutions by nontrivial configurations of nA magnetic fields
that are not associated with a Gauss law, thereby leading to an evasion of the no-hair conjecture.
Although some of the resulting nA solutions,  with $\Lambda<0$ and $\Lambda=0$, 
were already displayed in the literature \cite{Brihaye:2010wp}, 
\cite{Brihaye:2011nr}, we have presented here
 a unified framework, valid for both asymptotically flat and (anti)-de Sitter solutions.  
 In this context,
 new nA solutions approaching a dS background at infinity are presented and discussed.

 The solutions constructed here all have finite energy by virtue of the presence of the Chern-Simons
(CS) term, whose scaling properties are adequate for that purpose. In this sense, employing a CS term
is an $alternative$ to the use of higher order Yang-Mills curvature terms, irrespective of the
value of the cosmological constant \cite{Radu:2009rs}. For the dS case, a higher order YM curvature term was used 
 in \cite{Brihaye:2006xc} to obtain finite energy nA solutions.  
 The use of a CS term is
more restrictive than that of higher order YM curvatures, since in any given dimension there is a
unique (single trace) CS density. By contrast,
 the solutions in that case present a richer pattern.

Our study reveals that some of the features of the solutions 
are rather generic and independent of the presence of a cosmological term in the action.
In all cases, the nA black hole solutions emerge as perturbations of the RN solution, which becomes
unstable when embedded in a larger gauge group. 
Moreover, for any asymptotics, a set of particle like solutions is found for 
a critical value of the electric charge,   namely  only\footnote{
This feature can be understood by noticing that the cosmological constant does not 
enter the first integral (\ref{1SU2U1}).
Since (from regularity requirements) $w(0)=1$ and $\sigma(0)= \sigma_0>0$,
it follows that $Q=16\kappa$. Otherwise, $V(r)$ would doverge at the origin.
} for $Q=16\kappa$.

Following the approach in \cite{Brihaye:2011nr}, we could prove that some of our AdS solutions are also stable
with respect to first order of perturbations.
The equations for $\Lambda<0$ are rather similar to those given in Section 4 of \cite{Brihaye:2011nr}
for flat asymptotics, so we have not
 display  them here.
As expected, all stable solutions we have found have a nodeless magnetic potential $w(r)$. 
Physically, this means that the ballance between the attractive gravitational
force and the repulsive gauge $U(1)$
field is not affected when adding small enough non-Abelian fields.
The corresponding question of stability of dS solutions is complicated by the presence of a
cosmological horizon and we defer it to future work.
 
It may be interesting to contrast the situation with the
AdS$_5$ solutions of the pure EYM system  ($\kappa=0$) 
discussed in \cite{Manvelyan:2008sv,Ammon:2009xh}.
In both cases, we regard a $U(1)$ subgroup of the YM fields
as the gauge group of electromagnetics, and persuade the off diagonal gauge bosons to 
condense outside the horizon.
However, in the absence of a CS term, EYM black holes
with finite mass exist only in AdS spacetimes,   which are characterised by a Ricci flat horizon.
 Besides, such black hole solutions do not admit particle-like, globally regular limits.
In contrast, the EYMCS model discussed here admits also finite mass
AdS solutions with a Ricci flat horizon \cite{Zayas:2011dw},
 in addition to the solutions with a spherical horizon.
  
The EYM-$\Lambda$ black holes in  \cite{Manvelyan:2008sv}
describe, via gauge-gravity duality, a $p-$wave superfluid 
state in a strongly-coupled conformal field \cite{Ammon:2009xh}.
The possible relevance  of the EYM solutions  with a CS term 
in providing useful analogies 
to phenomena observed in condensed matter physics remains an interesting problem
to be understood in the future. At least in the AdS case, we expect our solutions to provide
duals of spherical superconductors. Further progress in this direction requires a computation of the conductivity as
a function of frequency by perturbing the YM fields, a problem
which is currently  under investigation.   It should be stressed that 
our results here
 signal the potentially important fact 
that some basic features exhibited by such 
solutions do not require the presence of a cosmological term in the action.
This may suggest that the solutions with flat or dS asymptotics may also be
useful in describing certain aspects of condensed matter systems.

 %%%%%%%%%%%%%%%%%%%%%%%%%%%%%%%%%%%%%%%%%%%%%%%%%%%%
\section*{Acknowledgements}
%%%%%%%%%%%%%%%%%%%%%%%%%%%%%%%%%%%%%%%%%%%%%%%%%%%%
 This work is carried out in the framework of Science Foundation Ireland (SFI) project
RFP07-330PHY. 
Y.B. is grateful to the
Belgian FNRS for financial support.
E.R. gratefully acknowledges support by the DFG.

\begin{small}

\end{small}

\end{document}